\documentclass[11pt]{article}
\usepackage{blois,epsfig}
\usepackage[latin1]{inputenc}
\usepackage[OT1]{fontenc}

\def\Journal#1#2#3#4{{#1} {\bf #2}, #3 (#4)}

\def\NIMA{{\em Nucl. Instrum. Methods} A}
\def\NPA{{\em Nucl. Phys.} A}
\def\EPJA{{\em Eur. Phys. J.} A}

\def\NPBS{{\em Nucl. Phys.} B {\em (Proc.Supp.)}}

\def\be{\begin{equation}}
\def\ee{\end{equation}}
\def\bea{\begin{eqnarray}}
\def\eea{\end{eqnarray}}

\begin{document}
\vspace*{4cm}
\title{SUPERNEMO: A NEXT GENERATION PROJECT TO SEARCH FOR
NEUTRINOLESS DOUBLE BETA DECAY}

\author{ Yu.A. SHITOV on behalf of the SuperNEMO Collaboration}

\address{Physics Department, Imperial College\\
Prince Consort Rd, London SW7 2BW, UK\\
$^*$E-mail: y.shitov@imperial.ac.uk}

\maketitle\abstracts{ The SuperNEMO project aims to search for
neutrinoless double beta decay ($0\nu\beta\beta$) up to a sensitivity
of 10$^{26}$ years for the $0\nu\beta\beta$ half-life (down to 50~meV
in the effective Majorana neutrino mass), using $\sim$100 kg of source
and a `tracko-calo' detector. The current status of the 2007--2009
R\&D programme is presented here, focusing on the most challenging
aspects: calorimetry, production of sources, low radioactivity
measurements, and tracker.}

\section{Introduction}

Discovery of neutrinoless double beta decay, $0\nu\beta\beta$:
$(A,Z) \rightarrow (A,Z+2) + 2e^-$, which is forbidden in the Standard Model
due to lepton number conservation, would prove the Majorana nature and
reveal new fundamental properties of the neutrino. The hierarchy and absolute
mass scale of the neutrino eigenstates would be determined in the case where
$0\nu\beta\beta$-decay is driven by light neutrino exchange, while new
physics could be tagged in the case of other possible
$0\nu\beta\beta$-decay mechanisms, such as
right-handed currents, R-parity violation of SUSY, etc~\cite{Suh05,Ver05}.

The principal difference in experimental techniques is whether the two
electrons emitted in the $\beta\beta$-decay are measured directly
(tracking + calorimetry or TPC) or not (geochemistry or calorimetry
only). Pure calorimeters (germanium semiconductors and
bolometers) are the $\beta\beta$-sources themselves and thus only
measure the total energy deposited by both electrons. 

In comparison with calorimeters, the direct methods have worse
efficiency and energy resolution, but better background rejection
and the possibility of measuring different isotopes. 
However, the most important feature is that the individual energies and
trajectories of both electrons can be measured. Obtaining this unique 
information is the only way to probe the decay mechanism once a
$0\nu\beta\beta$-signal has been found by any experiment.

The SuperNEMO project is the next step in direct experimental
$0\nu\beta\beta$-decay searches based on the `tracko-calo' technique
of the NEMO series of experiments, including the latest 
currently running NEMO-3 detector~\cite{NEMO3}.  
Inspired by the success of these experiments, the NEMO/ SuperNEMO 
Collaboration~\footnote{Includes $\sim$80
physicists from 12 countries (http://nemo.in2p3.fr).} has embarked on an
 R\&D programme (2007--2009) to design a detector with sensitivity 
down to 50~meV in the effective Majorana neutrino mass 
(up to 10$^{26}$ years for the $0\nu\beta\beta$ half-life) from measurements 
of $\sim$100 kg of source.

The SuperNEMO basic features and the current status of key R\&D
studies are presented in this article.

\section{Status of SuperNEMO R\&D}

The SuperNEMO project will extrapolate the NEMO-3 `tracko-calo'
technology to the new scale with the principal parameters shown in
Table~\ref{parameters}.

\begin{table}[hbt]
\caption{Comparison of the main NEMO-3 and SuperNEMO parameters.}
\label{parameters}
\begin{tabular}{|l|c|c|} \hline
Parameter & NEMO-3 & SuperNEMO  \\ \hline
Isotope & $^{100}$Mo & $^{150}$Nd or $^{82}$Se \\ \hline
Mass, kg    & $7$ & $100$--$200$ \\ \hline
Efficiency, \% & $18$ & $>30$ \\ \hline
Energy resolution at 1 MeV (3 MeV) e$^-$, FWHM in \% & $\sim 12$
($\sim 8$) & $\sim 7--8$ ($\sim 4$)\\ \hline
$^{208}$Tl in foil, $\mu$Bq/kg & $<20$ & $<2$ \\ \hline
$^{214}$Bi in foil, $\mu$Bq/kg & $<300$ & $<10$ (only for $^{82}$Se) \\ \hline
Internal background ($^{208}$Tl, $^{214}$Bi), counts/full mass/year & 0.5 & 0.5 \\ \hline
$T_{1/2}^{0\nu\beta\beta}$ sensitivity, $\cdot$10$^{26}$years & $>0.02$ & $>1$ \\ \hline
$<m_\nu>$ sensitivity, meV& $300$--$1300$ & $50$--$100$ \\ \hline
\end{tabular}
\end{table}

\subsection{Design}
\begin{figure}[bht]
\begin{center}
\raisebox{1.5cm}{\psfig{figure=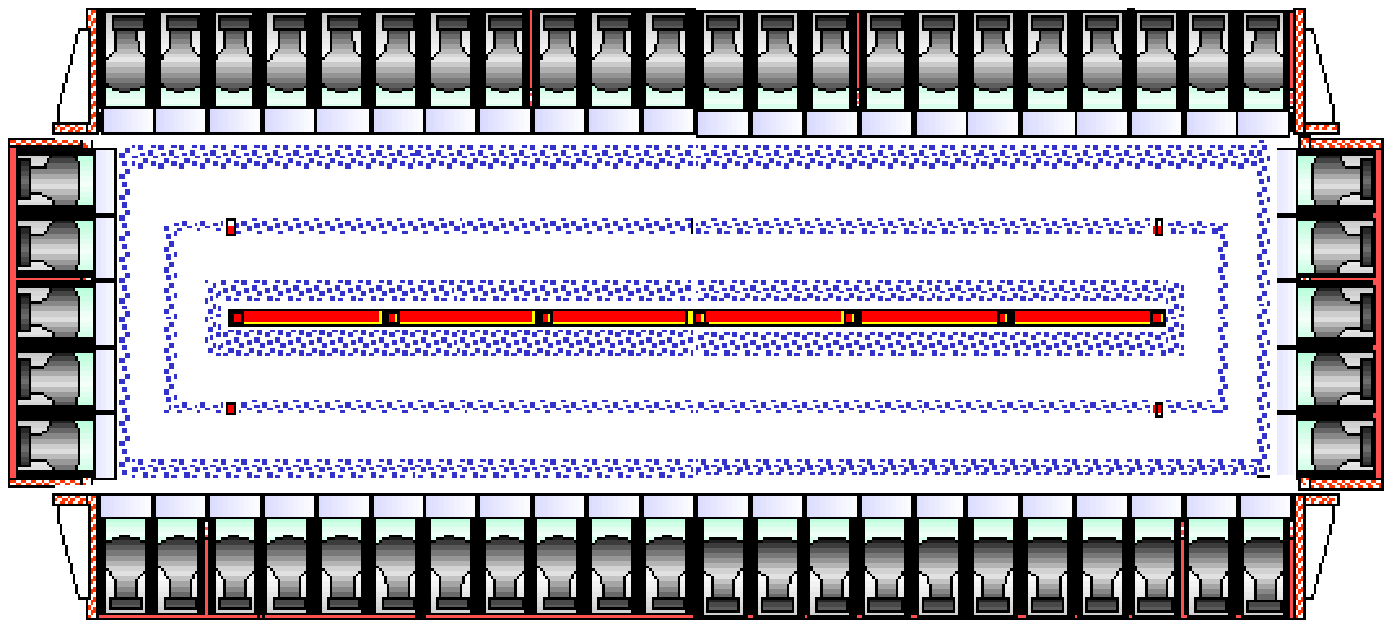,height=3.5cm,width=6.5cm}}
\hspace{1cm}
\psfig{figure=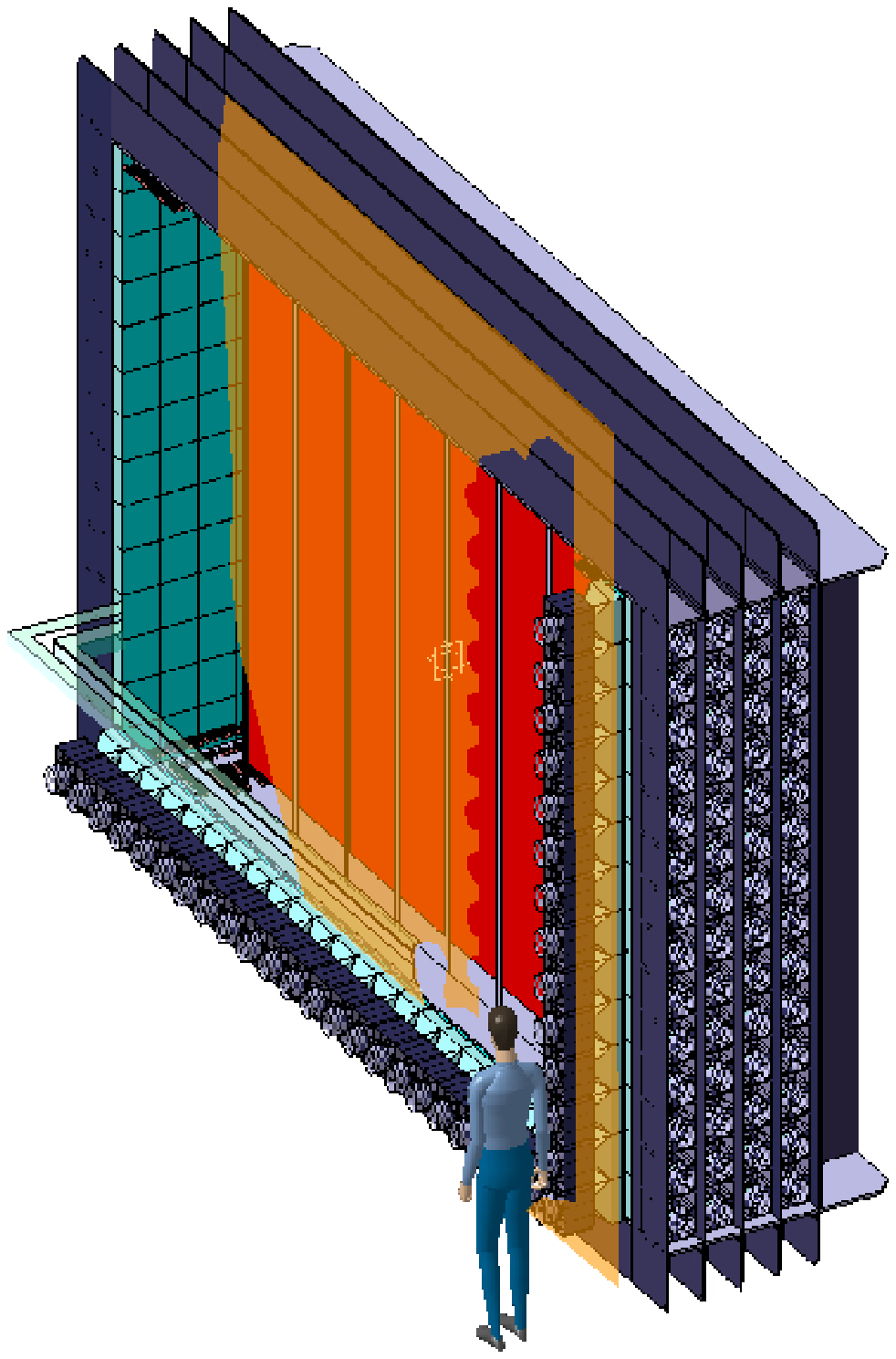,height=6.5cm,width=4.5cm}
\caption{Basic design of the SuperNEMO module: top (left) and 3D
  (right) views.}
\label{snemo}
\end{center}
\end{figure}

The SuperNEMO detector will follow a modular concept, consisting of $\sim$20
planar units with $\sim$5~kg of isotope per 5~$\times$~4~$\times$~1 m
module. The basic design of a module is shown in
Fig.~\ref{snemo}. Electrons emitted from a thin ($\sim$40 mg/cm$^2$)
$\beta\beta$-source foil in the middle of the module traverse a  
tracking chamber ($\sim$3000 wire drift cells operated in Geiger mode) 
before entering a calorimeter  (assembled from 1200 
20~$\times$~20~$\times$~2 cm
scintillators coupled with 8-inch PMTs). Options under study are
the use of long scintillator bars instead of blocks and an external gamma
calorimeter (veto) surrounding the module.

\subsection{Isotope Choice}
The physics criteria for the isotope choice are: i) large
$Q_{\beta\beta}$ to give a big phase space factor and better
background rejection, ii) large $T_{1/2}^{2\nu\beta\beta}$ to reduce
unavoidable $2\nu\beta\beta$-background, iii) large nuclear matrix
element (NME) to enhance the decay rate. Unfortunately, the latter is
rather unreliable as NME uncertainties remain quite large despite
recent progress in the development of calculation methods~\cite{Suh05}.

High natural isotope abundance, easy enrichment, radiopurification
and foil preparation are practical criteria required to produce 100~kg
of ultra-radiopure thin sources at a reasonable price. The main
candidates for SuperNEMO are $^{82}$Se and $^{150}$Nd. 

The full production chain is being studied now for selenium: i)
centrifugation enrichment has been tested~\footnote{Enrichment and
most of the purification have been funded by ILIAS
(http://www-ilias.cea.fr).}  producing 3.5~kg of $^{82}$Se; 100~kg
could be produced in 3 years; ii) purification by two methods
(chemical and distillation) has been carried out and checked at the kg scale;
iii) foil production has been redone with NEMO-3 technology and a new
technique is being tested.

Neodymium could be a promising isotope for measurements (less
background restrictions and possible physics benefits) but it requires
a more expensive enrichment procedure. The SuperNEMO Collaboration has
initiated studies with the aim of enriching $^{150}$Nd via the atomic
vapour laser isotope separation technique (AVLIS) with MENPHIS/CEA
facility.

\subsection{Calorimeter}

The energy resolution is a key factor in discriminating a 
$0\nu\beta\beta$-signal from $\sim$10$^5$--10$^6$ times as much unavoidable
$2\nu\beta\beta$-background. To reach a factor of two improvement
relative  to NEMO-3 (see Table~\ref{parameters}), with the
$\sim$1000~m$^2$ of detection surface in SuperNEMO is a
challenging task as the technology has already been well tuned over
many years. The search for the best design  includes: 
i) tests of different scintillator materials (plastic, liquid,
non-organic) produced by improved technology where possible; 
ii) maximisation of light collection, choosing optimal scintillator shape
and size, new and improved reflector coating materials; 
iii) development of new ultra-low background, high quantum efficiency (HQE)
PMTs, working closely with the Hamamatsu, Photonis and ETL companies; 
iv) design of a technical implementation of the calorimeter. Initially, the
required resolution was obtained for small (up to 6~$\times$~6
cm) detection surfaces; recently 8--8.5\% has been reached with
20~$\times$~20 cm plastic scintillator coupled with a new 8-inch HQE
(43\% in comparison with standard 25\%) Hamamatsu PMT. The final
decision on the calorimeter design is scheduled for the end of 2008.

\subsection{Source radiopurity}

The most dangerous internal sources are $^{208}$Tl and $^{214}$Bi
contaminations which must be reduced by factors 10 and 30, respectively, in
comparison with the NEMO-3 detector (see Table~\ref{parameters}).  Such
ultra-radiopurity is beyond the sensitivity of standard
low-background measurement techniques (1 kg~$\times$~month exposure
in $\sim$400~cm$^2$ HPGe). The dedicated {\it BiPo} detector~\footnote{
http://nemo.web.lal.in2p3.fr/BiPo/BiPo-webpage.htm} is being developed with
the aim of measuring $^{208}$Tl/$^{214}$Bi activities in thin foils
(12~m$^2$, 5 kg) at the level required, with one month of exposure,
by tagging the {\it Bi}smuth-{\it Po}lonium chain signature: an electron followed by a 
delayed alpha particle~\cite{BiPo}. The BiPo-I prototype has been built, tested 
and is now running in the Underground Laboratory of Modane (LSM,
France)~\footnote{http://www-lsm.in2p3.fr/}. Today,  18 BiPo-I capsules,
with a total detecting surface of $\sim$0.72~m$^2$, are measuring their own
background; the sensitivity recently reached is
A($^{208}$Tl)$<$7.5~$\mu$Bq/kg for a one month exposure in a 10~m$^2$
detector. In parallel, alternative BiPo-II and Phoswhich prototypes
have been built, tested, and are now being installed in the LSM. 
The first results of the sensitivity tests are expected at the 
end of this summer.

\subsection{Tracker}

Improvement of performance in tagging charged particles (e$^\pm$ and
$\alpha$) in the tracking chamber implies optimal choice of
construction materials, sizes of wires and cell, cell layout
(topology) design permitting automated wiring, working gas mixture,
and readout. Complex methodical tests with several small prototypes and
simulations have been done to develop the SuperNEMO tracker unit.  A
9-cell prototype has been built and has demonstrated required
performances with three different wire layouts. A 90-cell prototype is
in production now and will be ready this summer.  As an industrial
scale is required for assembling the SuperNEMO tracker ($\sim$500,000
wires must be processed), a dedicated wiring robot is being developed
for the mass production of drift cells.  The final decision on the
SuperNEMO tracker design is scheduled for the end of 2008.

\subsection{Miscellaneous}

\noindent {\it Simulations.} The SNSW~\footnote{SuperNemo SoftWare,
http://evalu29.uv.es/SuperNemoSW} package has been developed and first
simulations have proved that the SuperNEMO target sensitivity is, in 
principle, reachable with the parameters specified.

\noindent {\it Location.} It is initially planned that the BiPo
detector and up to the first five SuperNEMO modules will be hosted in
the Canfranc Underground
Laboratory~\footnote{http://ezpc00.unizar.es/lsc/index2.html}, while
the rest will be located in a new cavern at LSM which is expected to
be available in 2012.

\noindent {\it Schedule.} The first SuperNEMO module is planned to be
constructed in 2010-2011 with the full detector running in 2014 and
reaching the target sensitivity in 2016.

\section {Conclusion}

$0\nu\beta\beta$ studies have a potential of discovery to reveal
new fundamental properties of the neutrino  in particular and nature in
general. Several experiments with different techniques are required to
confirm definitely any possible signal observation.

Based on the successful experience of the NEMO detectors, the SuperNEMO 
R\&D programme is being carried out extensively and intensively, reaching 
the final phase and going forward to the TDR preparation in 2009.  
In terms of sensitivity and time scale, SuperNEMO is competitive with other
world-best $0\nu\beta\beta$-projects (see review~\cite{Pir06}); the
unique technique of the SuperNEMO detector could provide the possibility to
study the origin of $0\nu\beta\beta$-decay in the case of its
discovery.

\section*{Acknowledgements}
A portion of this work was supported by grant from RFBR
(No. 06-02-16672-a).

\end{document}